 \documentclass[aps,pre,preprint]{revtex4-1}

\pdfoutput=1

\usepackage{graphicx}
\usepackage{amsmath}
\usepackage{amsfonts}
\usepackage{wasysym}
\usepackage{tabularx}
\usepackage{color}

\begin{document}

\title{Stochastic anomaly and large Reynolds number limit in hydrodynamic turbulence models}

\author{A.A. Mailybaev}
\affiliation{Instituto Nacional de Matem\'atica Pura e Aplicada -- IMPA, Rio de Janeiro, Brazil}
\email{alexei@impa.br}

\begin{abstract}
In this work we address the open problem of high Reynolds number limit in hydrodynamic turbulence, which we modify by considering a vanishing random (instead of deterministic) viscosity. In this formulation, a small-scale noise propagates to large scales in an inverse cascade, which can be described using qualitative arguments of the Kolmogorov--Obukhov theory. We conjecture that the limit of the resulting probability distribution exists as $\mathrm{Re} \to \infty$, and the limiting flow at finite time remains stochastic even if  forcing, initial and boundary conditions are deterministic. This conjecture is confirmed numerically for the Sabra model of turbulence, where the solution is deterministic before and random immediately after a blowup. Then, we derive a purely inviscid problem formulation with a stochastic boundary condition imposed in the inertial interval. 
\end{abstract}

\maketitle

\section{Introduction}

The phenomenon of turbulence remains a major challenge in incompressible fluid dynamics. 
In the idealized formulation, when forcing is applied at a large scale $L$, the developed turbulence is considered to be statistically stationary, isotropic, homogeneous, and universal at small scales. 
The contemporary understanding of this phenomenon largely relies on the Kolmogorov-Obukhov theory~\cite{kolmogorov1941local,obukhov1941spectral}, which conjectures that 
the mean energy flux $\varepsilon$ from large to small scales together with the viscosity $\nu$ define all statistical properties of the small-scale dynamics. 
The problem of developed turbulence contains a natural dimensionless small parameter,  the inverse Reynolds number. Since the Reynolds number achieves very large values in applications, e.g., $\mathrm{Re} \sim 10^8$ in atmospheric flows, the theory accessing the limiting description as $\mathrm{Re} \to \infty$ would be the most appropriate. 

 The problem of describing the limit of large Reynolds numbers remains open. Mathematically, weak solutions of the Euler equations should be associated to the inviscid limit of the incompressible 
3D Navier--Stokes equations; however, extra admissibility conditions are required due to unphysical solutions~\cite{scheffer1993inviscid,shnirelman1997nonuniqueness,bardos2013mathematics}.  
A different approach involves a probabilistic description. 
The necessity of a stochastic approach for turbulent flows is widely accepted~\cite{frisch1999turbulence} and justified, e.g., by the exponential path separation controlled by positive Lyapunov exponents. However, the dynamical system viewpoint fails in the limit $\mathrm{Re} \to \infty$, since the Lyapunov exponents depend strongly on $\mathrm{Re}$. The stochastic formulation is traditionally achieved, e.g.,  by considering random large-scale forcing. Though the relevance of random forcing as a physical model can be questioned, it is a convenient theoretical assumption. 
On the other hand, the recent ideas of spontaneous stochasticity developed for Lagrangian trajectories in fully developed turbulent flows~\cite{bernard1998slow,eijnden2000generalized,falkovich2001particles,kupiainen2003nondeterministic} suggest the randomness as a consequence of non-uniqueness of solutions. Hence, the stochasticity may arise naturally in the equations, which look fully deterministic (deterministic forcing, initial and boundary conditions) as soon as the limit $\mathrm{Re} \to \infty$ is considered. 

This work is aimed to extend the ideas of spontaneous stochasticity, at the same time following the earlier studies on the influence of a small-scale noise~\cite{leith1972predictability,ruelle1979microscopic,eyink1996turbulence,eyink2014}.   
We provide a numerical evidence and a qualitative theoretical reasoning suggesting that the turbulent flow in the limit $\mathrm{Re} \to \infty$ is an intrinsically stochastic process well defined within a \textit{deterministic} formulation, 
i.e., for deterministic forcing, initial and boundary condition. Such a description is achieved by introducing an infinitesimal (vanishing as $\mathrm{Re}\to\infty$) \textit{random} perturbation of the viscosity, acting as a regularizing term in the governing equations. 
We show that this perturbation acts at the Kolmogorov viscous scale and induces an inverse cascade resulting in stochastic mixing at larger and larger scales~\cite{leith1972predictability}. 
This process can be described qualitatively using arguments of the Kolmogorov--Obukhov theory. As a result, the stochastic flow components persist at all scales in the limit of vanishing random viscosity.  

The conjecture that we call the \textit{stochastic anomaly} and verify numerically for the Sabra model of turbulence suggests the existence of a probability distribution in the limit $\mathrm{Re} \to \infty$, which is independent of the small-scale regularizing noise. Our numerical limiting solution is deterministic at times before the blowup, and it becomes random spontaneously at later times. On one hand, this means that the flow cannot be predicted exactly at a finite time, as opposed to the chaotic motion where a finite-time uncertainty can always be removed for sufficiently accurate initial conditions. On the order hand and more importantly, this shows that the limiting ($\mathrm{Re} \to \infty$) probability distribution is well-defined and, thus, can be predicted. We demonstrate theoretically and numerically that the limiting probability distribution satisfies the equations of fully inviscid dynamics with a proper stochastic boundary condition imposed in the inertial interval.

In Section~\ref{sec2}, we give a qualitative description of the inverse cascade of stochasticity, based on Kolmogorov arguments. Section~\ref{sec3} formulates the hypothesis of the stochastic anomaly in the limit of large Reynolds numbers, and Section~\ref{sec4} verifies this hypothesis numerically for the Sabra model of turbulence. Section~\ref{sec5} describes main features of the suggested stochastic description, as compared to the phenomena of deterministic chaos and spontaneous stochasticity of Lagrangian trajectories. Section~\ref{sec6} describes the small-scale statistics, which is used in Section~\ref{sec7} for implementing the purely inviscid formulation with a stochastic closure at small scales. Section~\ref{sec8} provides numerical tests of this probabilistic formulation. We finish with some conclusions. 

\section{The inverse cascade of stochasticity}
\label{sec2}

The importance of small-scale noise for describing a turbulent flow was recognized for long time, see e.g.~\cite{leith1972predictability,ruelle1979microscopic,eyink1996turbulence}. Due to large separation between the scale, where the noise is ``injected'', and its eventual observation at large scales of the flow, this becomes a cascade phenomenon. As a result, the noise amplification is governed by a power-law, as opposed to the exponential separation of trajectories in finite-dimensional deterministic chaos. In this section, we provide a qualitative argument based on the Kolmogorov approach describing the propagation of stochastic perturbations from small to large scales. 

The Kolmogorov-Obukhov theory~\cite{kolmogorov1941local,obukhov1941spectral} conjectures that 
the mean energy flux $\varepsilon$ from large to small scales together with the viscosity $\nu$ define all statistical properties of the small-scale dynamics. At scales of the inertial interval, $\eta \ll \ell \ll L$, where 
\begin{equation}
\eta = \nu^{3/4}\varepsilon^{-1/4}
\label{eq_nu}
\end{equation}
is the Kolmogorov length, both forcing and viscosity can be neglected.
The dimensional analysis 
yields power-laws for the moments of velocity fluctuations, 
$\langle|\delta v|^p\rangle\sim \varepsilon^{p/3} \ell^{p/3}$, see e.g.~\cite{landau2013fluid}. 
These scaling laws are only approximate due to the existence of anomalous corrections in the exponents~\cite{frisch1999turbulence}, which still lack the comprehensive theoretical explanation. 
At smaller scales, $\ell \ll \eta$, the flow is dominated by viscosity. Choosing the large-scale speed as $V = \varepsilon^{1/3} L^{1/3}$, the Reynolds number is related to the inertial interval span as $\mathrm{Re} = VL/
\nu = (L/\eta)^{4/3}$. 

Assuming locality of nonlinear interactions, the energy produced by forcing at scale $L$ is transferred successively to smaller and smaller scales, until it dissipates at the smallest viscous scales. 
In the inertial range, a turnover time  at scale $\ell$ is given by $\tau \sim \varepsilon^{-1/3} \ell^{2/3}$, and it is a characteristic time for the energy transfer to the smaller scale $\ell/2$. Hence, a total time for the energy transport from the forcing to viscous scales is given by a convergent geometric series, independent of viscosity. This observation constitutes the dissipation anomaly due to Onsager~\cite{onsager1949statistical,eyink2006onsager}: the energy flux to small scales remains finite for vanishing viscosity, i.e., in the limit of high Reynolds numbers, $\mathrm{Re} \to \infty$. 

A similar phenomenon occurs for a noise, but with an opposite direction from small to large scales~\cite{leith1972predictability,eyink1996turbulence}. There are various physical reasons for introducing a noise in turbulent flow at small scales, e.g., thermodynamic molecular motion or parameter uncertainties. Apparently,  an effect of this noise at large scales is independent on its particular form and origin. In this work, we model the noise by considering an uncertainty for the viscosity parameter $(1+x)\nu$, where $x > -1$ is a random number that describes a relative viscosity deviation. 
Such an uncertainty may result from various physical properties: humidity fluctuations in air or temperature  variations in water, etc. The measurement and numerical errors also serve as a justification. This viscosity perturbation yields an additional viscous term $x\nu \Delta\mathbf{v}$ in the governing Navier--Stokes  equations. In the inertial interval, one can use the Kolmogorov estimate $\delta v\sim \varepsilon^{1/3} \ell^{1/3}$ for the speed variation and $\ell^{-2}$ for the Laplace operator, which leads to the relation 
	\begin{equation}
	x\nu \Delta\mathbf{v} \sim \sigma_x\nu \ell^{-5/3}\varepsilon^{1/3}
	= \sigma_x \left(\frac{\eta}{\ell}\right)^{4/3} \ell^{-1/3}\varepsilon^{2/3},
	\label{eq1}
	\end{equation}
where we used Eq.~(\ref{eq_nu}) and denoted by $\sigma_x$ the standard deviation of the relative viscosity perturbation $x$.
Similarly, the quadratic convective term $(\mathbf{v}\cdot \nabla)\mathbf{v}$ has the scaling $\ell^{-1/3}\varepsilon^{2/3}$. 
This shows that, for $\sigma_x \lesssim 1$, the random viscous term can be neglected at scales of the inertial interval, $\ell \gg \eta$, just like one neglects the deterministic viscosity. Thus, the random viscous term becomes important only at the Kolmogorov scale, $\ell \sim \eta$. 
This can be interpreted as the ``injection'' of a stochastic component to the flow at a small-scale end of the inertial interval.  

The effect of random viscosity perturbation on the dynamics at Kolmogorov scale can be estimated from the Newton's law, i.e., 
comparing a random viscous force in Eq.~(\ref{eq1}) at $\ell = \eta$ with the acceleration term $D\mathbf{v}/Dt$. The stochastic velocity component grows in magnitude to the mean value $\langle|\delta v|\rangle \sim \varepsilon^{1/3} \eta^{1/3}$ in the characteristic time
	\begin{equation}
	t_K \lesssim \frac{\langle|\delta v|\rangle}{\sigma_x\eta^{-1/3}\varepsilon^{2/3}} 
	= \frac{\eta^{2/3}}{\sigma_x\varepsilon^{1/3}}.
	\label{eq2}
	\end{equation}
Considering the large-scale characteristic time as $T = L/V = \varepsilon^{-1/3}L^{2/3}$ yields
	\begin{equation}
	\frac{t_K}{T} \lesssim \frac{1}{\sigma_x} \left(\frac{\eta}{L}\right)^{2/3} = \frac{1}{\sigma_x\sqrt{\mathrm{Re}}}.
	\label{eq3}
	\end{equation}
This means that the time of development of stochastic dynamics at the Kolmogorov scale is negligible for the large-scale flow if 
	\begin{equation}
	\sigma_x \gg \frac{1}{\sqrt{\mathrm{Re}}}.
	\label{eq4}
	\end{equation}
In the limit $\mathrm{Re} \to \infty$, Eq.~(\ref{eq4}) ensures that even a vanishingly small \textit{relative} perturbation of the viscosity is sufficient for our further arguments. We note that Eq.~(\ref{eq2}) yields a rather mild estimate based on the linear growth of disturbances. Similar arguments taking into account the exponential path separation yield much smaller values of $t_K$~\cite{ruelle1979microscopic}.  

Injection of a random flow component at small scales of the inertial interval leads to its turbulent transport towards the largest scale $L$ due to nonlinear interaction~\cite{leith1972predictability,eyink1996turbulence}, 
i.e., to an inverse cascade of stochasticity. In this process, the time required for developing the stochastic velocity component at scale $2\ell$ due to analogous component at scale $\ell$ is comparable to the turnover time $\tau \sim \varepsilon^{-1/3} \ell^{2/3}$, which rapidly increases with the transition to larger and larger scales. Because of the power-law relation for $\tau$, the total time of the transition through the inertial interval is given by a convergent geometric series, i.e., it remains finite in the limit $\mathrm{Re}\to\infty$ and does not dependent on viscosity~\cite{eyink2014}. Also, due to the increase of turnover times for larger scales, we expect that the fast small-scale stochastic dynamics is self-averaged on top of the slow large-scale flow, losing the dependence on a probability distribution of the random variable $x$.  

\section{Stochastic anomaly}
\label{sec3}

Reynolds numbers appearing in applications of the developed turbulence attain very large values. For example, in typical atmospheric flows, $\mathrm{Re} \sim 10^8$ results in the huge scale separation between the viscous and large-scale motion, $L/\eta \sim \mathrm{Re}^{3/4} \sim 10^6$. Since the large-scale flow is usually of primary interest, it is reasonable to consider the limit of high Reynolds numbers. In this section we formulate a hypothesis of the stochastic anomaly, which is a consequence of the inverse cascade of stochasticity considered in the limit $\mathrm{Re} \to \infty$. This hypothesis will be confirmed numerically for the Sabra shell model of turbulence in the following sections. 

The two properties of the inverse cascade of stochasticity are important for taking the limit $\mathrm{Re} \to \infty$. First, we argued that specific properties of the small-scale noise are ``forgotten'' at large scales, due to an increasing turnover time. Second, we saw that the arrival time of a stochastic perturbation to a specific scale has a finite limit for high $\mathrm{Re}$. This suggests that the limit $\mathrm{Re} \to \infty$ should be defined in terms of probability distributions for the velocity field. In this definition, the random viscosity plays a role of a regularization term, similarly to the role of a deterministic viscosity in the Burgers equation leading to a shock wave in the inviscid limit.

Therefore, our conjecture states that, in the limit of high Reynolds numbers formulated for a vanishing random viscosity, a limiting flow is described by a time-dependent probability distribution determined for given deterministic forcing, initial and boundary conditions.
Since this limiting solution is stochastic, despite all the conditions of the flow are deterministic, we call it the stochastic anomaly. 

\section{Numerical simulations for Sabra model}
\label{sec4}

Direct numerical simulations of the incompressible 3D Navier--Stokes equations, which accurately resolve all scales involved in the dynamics, are limited to rather low Reynolds numbers due to computer limitations. 
Still, condition (\ref{eq4}) shows that verification of the stochastic anomaly is possible at the boundary of its validity. 
For example, $\mathrm{Re} = 100$ requires $\sigma_x \gg 0.1$, which is satisfied for $\sigma_x \sim 1$ when the random viscosity perturbation $x\nu$ remains small. However, computing the probability distribution  requires a large number of simulations, each for a different random value of $x$, leading to extremelly high requirements for computation resources. 

Instead, in this work we perform the detailed verification using the Sabra model of turbulence~\cite{l1998improved}, which is a natural playground for the studies on developed turbulence. This model allows accurate numerical simulations for very 
high Reynolds numbers and possesses nontrivial properties of the Kolmogorov--Obukhov theory: energy cascade, dissipation anomaly, scaling power laws for velocity moments 
and anomalous corrections close to the ones for the Navier--Stokes equations. 
The Sabra model is obtained by reducing dynamics to a discrete sequence of shells $|\mathbf{k}| = k_n$ in the Fourier space for the geometric 
progression of wavenumbers $k_n = k_0\lambda^n$, $n = 1,2,3,\ldots$ (we use $k_0 = 1$ and $\lambda = 2$). 
The turbulent ``flow'' is described by complex velocity variables $u_n(t)$, which mimic the velocities at the corresponding shells. Thus, the velocity $u_n$ characterizes the scale $\ell \sim k_n^{-1}$. The model equations are
	\begin{equation}
	\begin{array}{rcl}
	\dot{u}_n 
	& = & \displaystyle
	ik_n\left(2u_{n+2}u_{n+1}^*-\frac{1}{2}u_{n+1}u_{n-1}^*
	+\frac{1}{4}u_{n-1}u_{n-2}\right)
	\\[12pt] && \displaystyle
	-(1+x)\nu k_n^2u_n+f_n, \quad n = 1,2,3,\ldots,
	\end{array}
	\label{eq5}
	\end{equation}
where $f_n$ is the forcing and $(1+x)\nu$ is the viscosity with a random perturbation; the boundary conditions are chosen as $u_{0} = u_{-1} = 0$. Eq.~(\ref{eq5}) has the inviscid invariants: energy $E = 
\sum |u_n|^2$ and helicity $H = \sum (-1)^nk_n|u_n|^2$. 

In our simulations, we choose the large-scale deterministic non-stationary forcing $f_2 = e^{2it}$ and $f_3 = 1-t$ and initial conditions $u_1 = 1/2+i$, $u_2 = 1$ at $t = 0$ 
(all other components are zero)  
in the model with 40 shells. 
In this case, the characteristic values $L$, $V$ and $T$ are of order 1 (a.u.), and the Reynolds number is  
defined as $\mathrm{Re} = 1/\nu$. The values up to $\mathrm{Re} = 10^{10}$ are considered. In each test, numerical simulations 
are performed for $10^4$ different values of the viscosity $(1+x)\nu$, where $x$ (fixed for each simulation) is taken as a random variable 
uniformly distributed in the interval $|x| \le \sqrt{3}\,\sigma_x$ with $\sigma_x = 0.3$. Additional tests with different initial conditions, 
forcing and distributions for $x$ led to the same conclusions. 

\begin{figure}
\centering
\includegraphics[width = 0.9\textwidth]{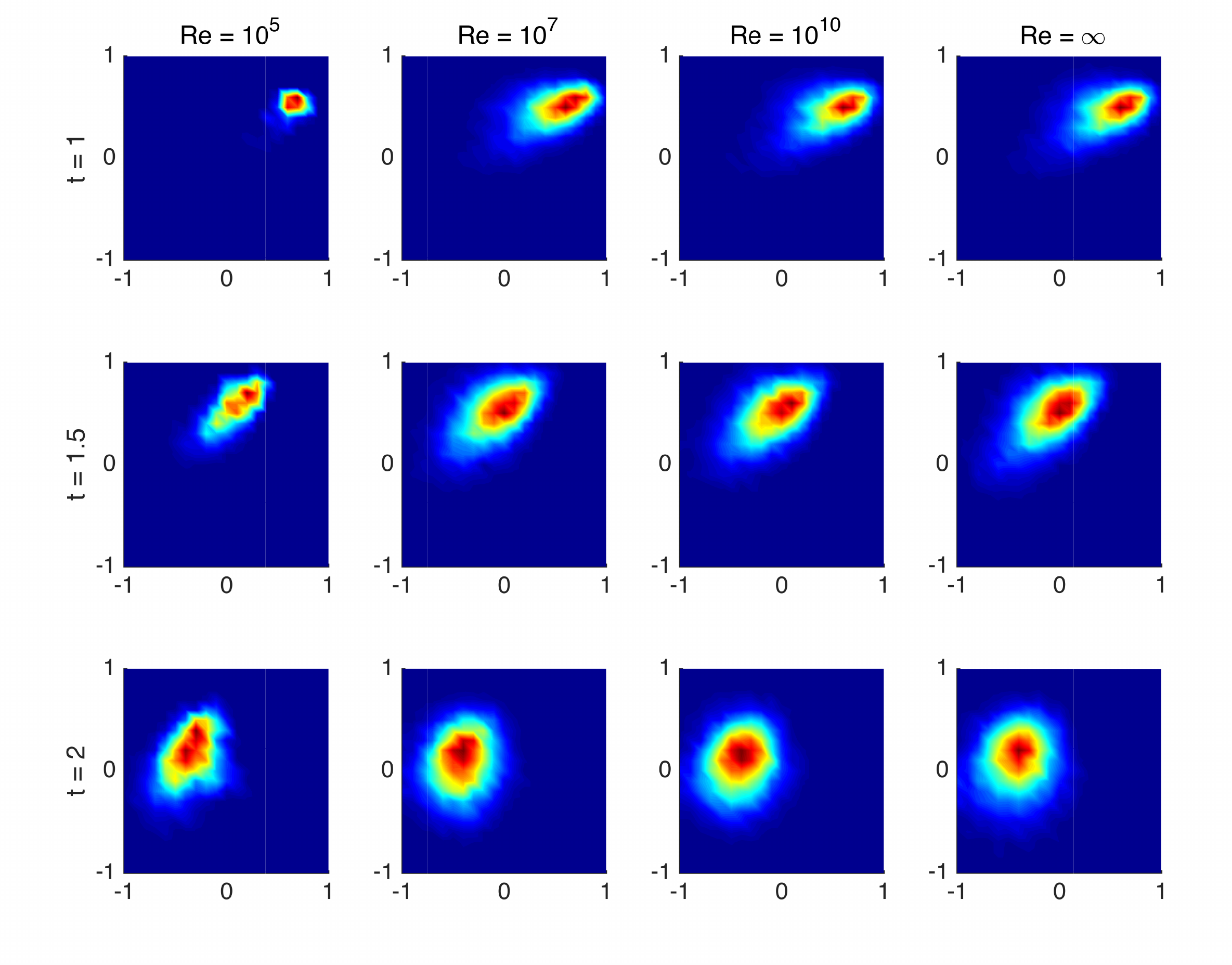}
\caption{Probability density, from blue (zero probability) to red (maximum probability) of the velocity $u_2(t)$ on complex plane at different times (rows) and Reynolds numbers (columns), obtained for deterministic initial conditions and forcing. The last column is obtained using the inviscid stochastic formulation of Section~\ref{sec7} with only 7 shells.}
\label{fig1}
\end{figure}

\begin{figure}
\centering
\includegraphics[width = 1\textwidth]{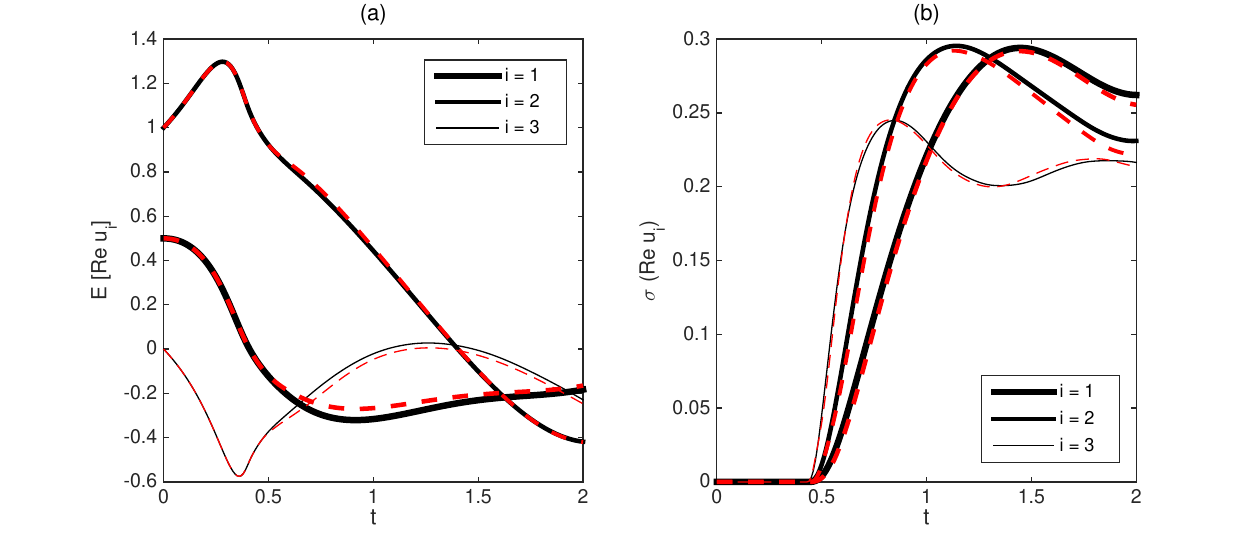}
\caption{(a) Mean values and (b) standard deviations for real parts of the large-scale shell speeds $u_1$, $u_2$ and $u_3$ as functions of time. The speeds become stochastic after the blowup at $t = 0.42$. Solid black curves correspond to  $\mathrm{Re} = 10^{10}$, and dashed red curves to $\mathrm{Re} = 10^7$.}
\label{fig1extra}
\end{figure}

Fig.~\ref{fig1} (the first three columns) presents the numerical probability densities for the large-scale velocity $u_2(t)$ at different times and for different Reynolds numbers. 
One can clearly see the stochastic form of large-scale dynamic, which converges with the increase of $\mathrm{Re}$. This confirms our hypothesis of the stochastic anomaly suggesting that, for high Reynolds numbers, the flow remains stochastic with a well-defined limiting probability distribution. Note, however, that the convergence is rather slow: one can clearly recognize the limiting pattern for $\mathrm{Re} = 10^7$, while the distortion is still rather strong for $\mathrm{Re} = 10^5$.  
In Fig.~\ref{fig1extra} we compare mean values and standard deviations of the large-scale speeds for $\mathrm{Re} = 10^7$ and $10^{10}$. 
We stress again that the presented probability distributions correspond to specific deterministic initial and forcing conditions. 

Fig.~\ref{fig1extra}(b) shows that the stochastic components of the flow do not exist up to the time $t_b \approx 0.42$. This time represents the blowup in the inviscid Sabra model, when the enstrophy $\sum k_n^2|u_n|^2$  grows to infinity and velocities at all scales get a power-law excitation~\cite{dombre1998intermittency,mailybaev2012renormalization}.  
Such behavior fully agrees with the regularity result~\cite{constantin2007regularity} proving uniqueness of the solution for the inviscid Sabra model at times before the blowup. Nonzero standard deviations for $t > t_b$ identify the blowup as a source of the spontaneous stochasticity: the solution is deterministic for  $t \le t_b$ and random for $t > t_b$.

In this section we presented the large-scale shell speeds, whose behavior is the most important from practical point of view. However, the nature of turbulent flow is in small-scale fluctuations~\cite{frisch1999turbulence,landau2013fluid}. We return to the study of the probability distribution for small-scale velocities in Section~\ref{sec6}.  

\section{Stochastic anomaly vs. deterministic chaos and spontaneous stochasticity of particles}
\label{sec5}

It is common to think that the developed turbulence cannot be described in terms of the theory of deterministic chaos. The spontaneous stochasticity of Lagrangian trajectories~\cite{bernard1998slow,eijnden2000generalized,falkovich2001particles,kupiainen2003nondeterministic} is one of the properties that highlights the conceptual difference. In this section we discuss these issues from the point of view of the stochastic anomaly. 

In finite-dimensional deterministic chaos, the concept of large-time unpredictability is based on exponential separation of solutions, which start at close initial points. A rate of separation is determined by positive Lyapunov exponents. However, the system dynamics is deterministic at every finite time: sufficiently (exponentially) close initial conditions yield a finite-time prediction with an arbitrary accuracy. The probabilistic description is introduced by considering a chaotic attractor in the limit $t \to \infty$, i.e., an invariant probability measure. This stochastic approach is applicable at sufficiently large times describing a statistical equilibrium. 
Such properties clearly distinguish the deterministic chaos from the stochastic anomaly: The limiting ($\mathrm{Re} \to \infty$) turbulent flow is truly stochastic at finite times, even if the initial conditions are fully deterministic. Furthermore, its probability distribution is time-dependent, not stationary. 

\begin{figure}
\centering
\includegraphics[width = 0.5\textwidth]{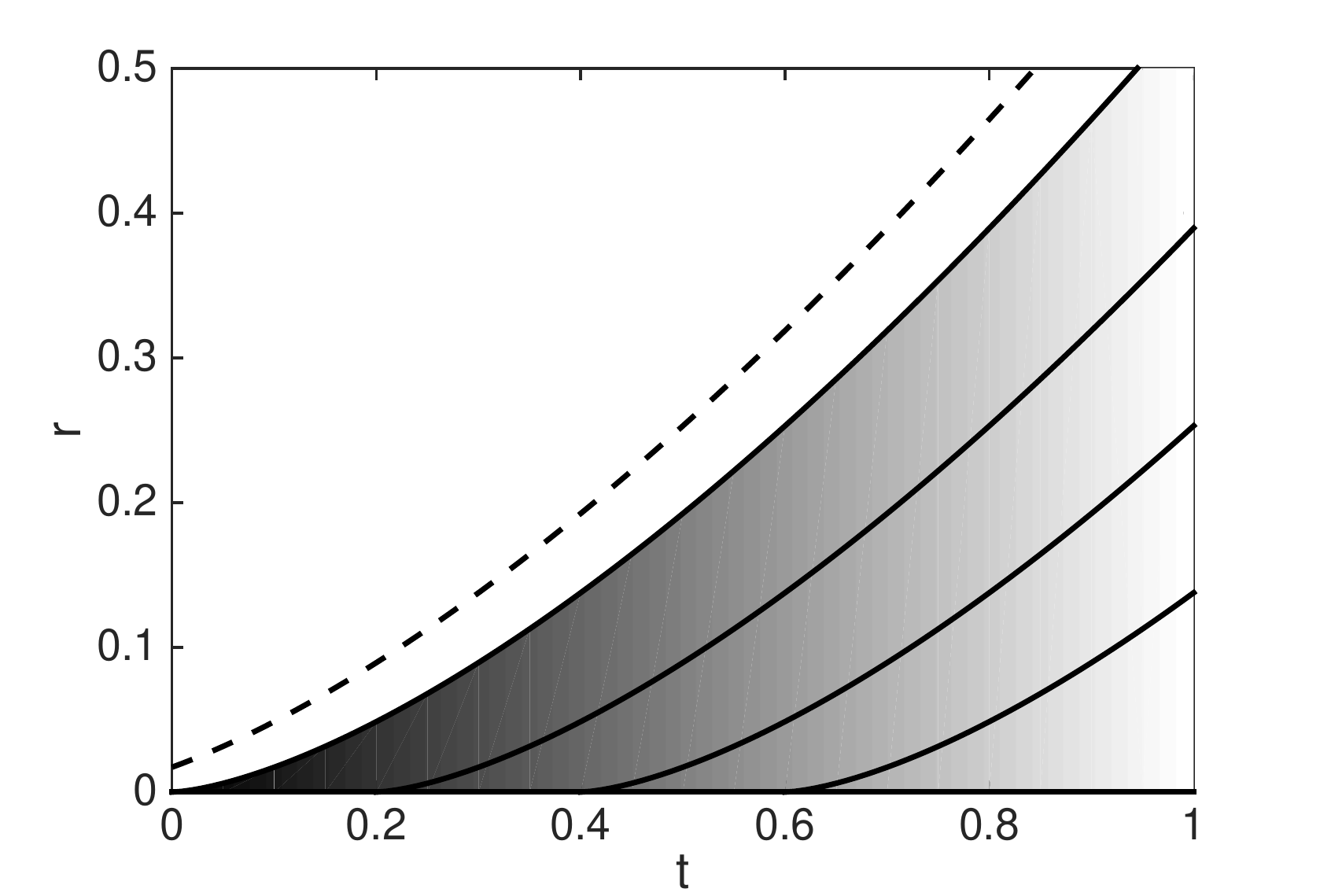}
\caption{Non-unique solutions of Eq.~(\ref{eqA1}) with the initial condition $r(0) = 0$ (solid lines) span a gray region, which grows as a power-law with time. The dashed line shows a solution with a small non-zero initial condition.}
\label{fig_ex}
\end{figure}

In order to see how this is possible, let us consider a simple ordinary differential equation
	\begin{equation}
	\dot r = r^{1/3},
	\label{eqA1}
	\end{equation}
which mimics a particle position $r$ at time $t$ with the initial condition $r(0) = 0$. Here the velocity $v = r^{1/3}$ is chosen such that it satisfies the Kolmogorov scaling. There is a family of solutions 
	\begin{equation}
	r(t) = \left\{\begin{array}{ll}
	0 & t \le t_s; \\[7pt]
	\left(\frac{2(t-t_s)}{3}\right)^{3/2}, & t > t_s ;
	\end{array}\right.
	\label{eqA2}
	\end{equation}
where $t_s$ is an arbitrary parameter denoting a spontaneous time, when the particle starts moving, see Fig.~\ref{fig_ex}. This example shows the non-uniqueness of the trajectories, inherent in Kolmogorov scaling laws. For Eq.~(\ref{eqA1}), this non-uniqueness is the well-known fact in differential equations, because the right-hand side if not Lipschitz continuous. For turbulent flows, these ideas appeared and were further elaborated for fluid particle trajectories assuming given (fixed or stochastic) rough velocity fields, see e.g.~\cite{bernard1998slow,eijnden2000generalized,falkovich2001particles,kupiainen2003nondeterministic}. As one can see from Eq.~(\ref{eqA2}), a separation between two solutions with close initial conditions grows as power-law, not exponentially. Moreover, if one of the initial conditions is at the origin, different separations can be achieved at a given time $t$ by choosing different parameters $t_s$. This shows how the stochasticity emerges instantaneously due to the non-uniqueness, as opposed to the exponential path separation in the deterministic chaos. Note also that the spontaneous stochasticity does not require the system to be chaotic, as follows from the above example and can be observed in the Gledzer shell model of turbulence~\cite{mailybaev2015spontaneous}.
  
Both in Eq.~(\ref{eqA1}) and in the theory describing non-unique particle trajectories~\cite{bernard1998slow,eijnden2000generalized,falkovich2001particles,kupiainen2003nondeterministic}, a singularity is introduced explicitly in the governing equations. A more sophisticated process drives the stochastic  anomaly as described in Section~\ref{sec4}. Here, equations (\ref{eq5}) do not feature any singularity in the right-hand side. Instead, a singularity leading to the non-uniqueness appears in the solution itself at the finite-time blowup.

Though the unpredictability of spontaneously stochastic turbulent flow is qualitatively different from the unpredictability in deterministic chaos, both lead to similar practical conclusions on essential limitations for finite-time predictions. The stochastic anomaly, however, comes along with a solution to this problem: it suggests that the regular probability distribution exists in the limit $\mathrm{Re} \to \infty$. In this sense, the spontaneous stochasticity is a property already inherent in the inviscid flow equations, i.e., inviscid Sabra model or incompressible Euler equations. This makes the limiting probability distribution a true physical solution of a ``deterministic'' inviscid problem that can be computed as a function of time and, thus, accurately predicted. In the following sections we suggest how this can be done in the framework of the Sabra shell model.

\section{Small-scale behavior}
\label{sec6}

\begin{figure}
\centering
\includegraphics[width = 0.98\textwidth]{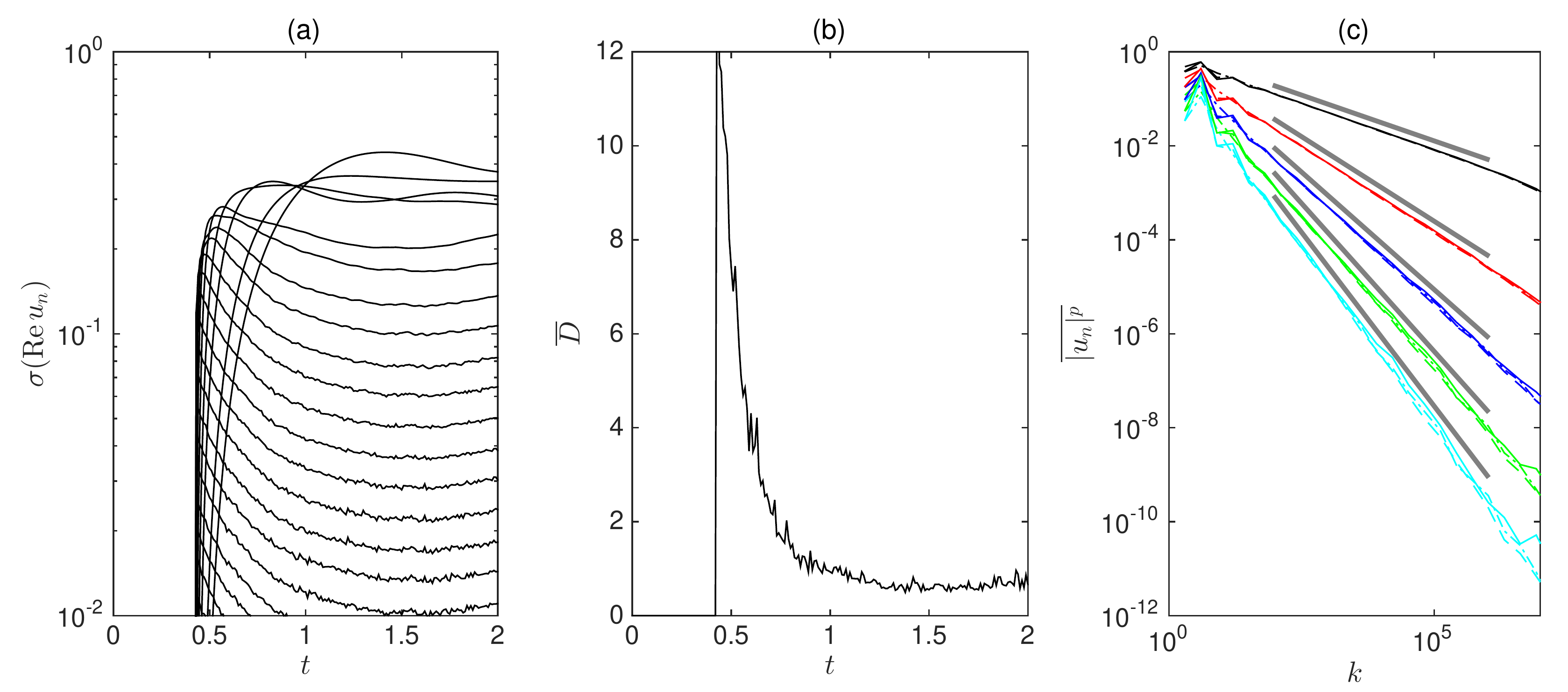}
\caption{(a) Standard deviations of real parts of shell velocities (logarithmic vertical scale); at last time, the shell numbers $n = 1,2,\ldots$ are ordered from top to bottom. (b) Instantaneous average of the dissipation rate $D = 2(1+x)\nu \sum k_n^2|u_n|^2$.  (c) Velocity moments $\overline{|u_n(t)|^p}$ at times $t = 1,\,1.5,\,2$ (solid, dashed and dash-dot lines) for $p = 1,2,\ldots,5$ (black, red, blue, etc.). The graphs are based on  
the solutions with $\mathrm{Re} = 10^{10}$, see Section~\ref{sec4}. 
}
\label{fig2}
\end{figure}

In this section, we focus on the time-dependent statistics at small-scales, i.e., in the inertial interval, for the simulations of Section~\ref{sec4}. Fig.~\ref{fig2}(a) shows the standard deviations for real parts of the shell speeds. One can see that all these quantities become non-zero after the blowup and change continuously with time (a small noise on top of the curves has a numerical origin due to finite sampling). The forcing range of shells, $n \lesssim 4$ at later times, can be clearly distinguished. In the inertial interval corresponding roughly to the shells $n \gtrsim 5$, the standard deviations accompany a decrease of the mean dissipation rate in time followed by a small increase, as one can see in Fig.~\ref{fig2}(b). Recall that our simulations use non-stationary deterministic forcing and, hence, no stationarity is expected in Fig.~\ref{fig2}(a,b). 

For stationary turbulence, the scaling of time-averaged velocity moments $\langle|u_n|^p\rangle \sim k_n^{-\zeta_p}$ is known to be anomalous, 
deviating from the Kolmogorov prediction $\zeta_p = p/3$~\cite{biferale2003shell}. This reflects the intermittency phenomenon, characterized by strong short-time bursts of shell speeds separated by regions of slow behavior. 
In our simulations, the probability distribution is non-stationary, changing with time at all scales.
Considering the averaging at a fixed time denoted with a bar (instead of the common time averaging), we obtain an important, though expected, observation that the scaling laws
are valid instantaneously with the same exponents, see Fig.~\ref{fig2}(c). 
In the figure, the graphs at times $t = 1,$ $1.5$ and $2$ are presented in the logarithmic scale and have the same slope for each $p$ 
(vertical shifts are used for better comparison). 
These slopes shown by gray lines are given by the anomalous exponents $\zeta_1 = 0.39$, $\zeta_2 = 0.72$, $\zeta_3 = 1$, $\zeta_4 = 1.26$, $\zeta_5 = 1.48$ of 
the stationary turbulence~\cite{l1998improved}. 

Our numerical observations suggest that the stationarity assumption for the description of the inertial interval can be relaxed, if one averages over the instantaneous (spontaneously stochastic) distribution, instead of performing the time averaging. Furthermore, this distribution depends slowly with time (except in a vicinity of the blowup). Thus, the intermittency remains a property of a single realization of the stochastic process, while the time-dependence of the probability distribution at all scales is characterized by the large-scale time $T$ determined by the  forcing or initial/boundary conditions. 

\begin{figure}
\centering
\includegraphics[width = 0.7\textwidth]{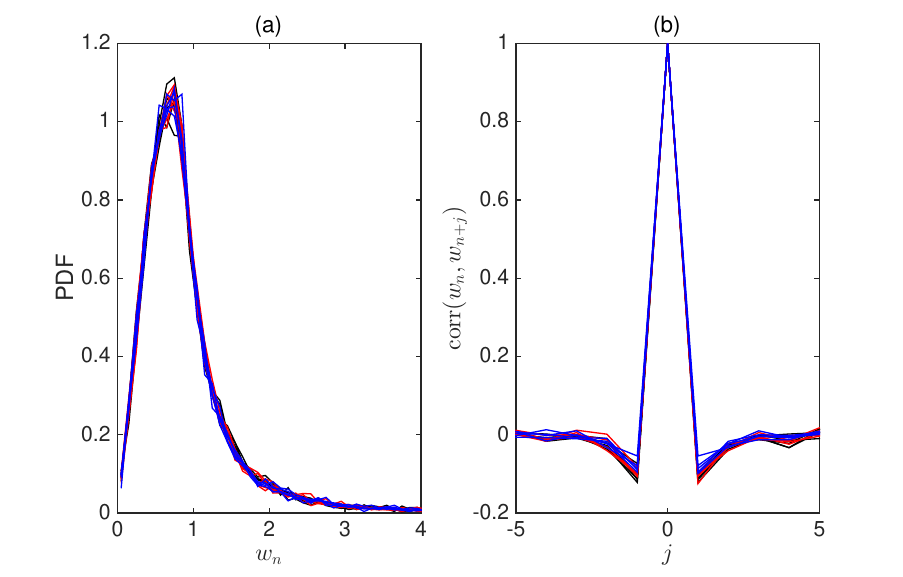}
\caption{(a) Probability density function for the shell speed multipliers $w_n = |u_n/u_{n-1}|$. (b) Correlation coefficients $\mathrm{corr}(w_n,w_{n+j})$. The results are shown at times $t = 1,\,1.5,\,2$ (green, red and blue) with $n = 12,\ldots,15$. Collapse of all the curves confirms universality of the probability distribution for velocity multipliers at small scales. 
}
\label{fig4}
\end{figure}

A deeper understanding is achieved by considering the 
Kolmogorov hypothesis~\cite{kolmogorov1962refinement,chen2003kolmogorov} on the universality of statistics of velocity multipliers studied in the context of stationary 
turbulence. For the Sabra model, this hypothesis reduces to considering the absolute ratios $w_n$ and the phases $\Delta_n$ defined as~\cite{benzi1993intermittency,eyink2003gibbsian}. 
	\begin{equation}
	w_n = |u_n/u_{n-1}|,\quad \Delta_n = \arg(u_{n-2}u_{n-1}u_n^*).
	\label{eqB1}
	\end{equation}
The two variables $\mathbf{z}_n = (w_n,\Delta_n)$ determine multiplicatively the shell speed $u_n$ for the known speeds $u_{n-1}$ and $u_{n-2}$ as
	\begin{equation}
	u_n = u_{n-1}w_ne^{-i\Delta_n+i\theta_{n-2}},\quad \theta_{n-2} = \arg u_{n-2}.
	\label{eqB1b}
	\end{equation}
It is conjectured~\cite{eyink2003gibbsian} that the probability distribution of the multiplier variables 
	\begin{equation}
	P(\ldots,\mathbf{z}_{n-1},\mathbf{z}_{n},\mathbf{z}_{n+1},\ldots)
	\label{eqB2}
	\end{equation}
is universal in the inertial interval, which is confirmed by numerical simulations. Furthermore, the distribution (\ref{eqB2}) is short-range (correlations between $\mathbf{z}_n$ and $\mathbf{z}_{n+j}$ decay rapidly with increasing $|j|$) and it is homogeneous in $n$ (invariant under the shift $\mathbf{z}_n \mapsto \mathbf{z}_{n+j}$ for any $j$). Our numerical simulations lead to the same conclusions, see Fig.~\ref{fig4}.

The universality demonstrated in Fig.~\ref{fig4} (also confirmed for the phases $\Delta_n$) is, in fact, a stronger result, because it is verified at every time for the instantaneous non-stationary probability distribution, i.e., not in the stochastic equilibrium. This means that the spontaneously stochastic distribution of the Sabra model, written for the multipliers, has a universal form at the side of small scales. This can be stated in a different way: the non-stationary probability distribution (defined in the limit $\mathrm{Re}\to\infty)$ evolves in time under the small-scale boundary condition given by the universal multipliers distribution (\ref{eqB2}). We now show how this boundary condition can be implemented in order reduce the problem to a purely inviscid evolution of the probability distribution.  

\section{Stochastic small-scale boundary condition for inviscid turbulent flow}
\label{sec7}

The evolution of a statistical distribution $P_n(u_1,\ldots,u_n)$ for the purely inviscid Sabra model is governed by the continuity equation
	\begin{equation}
	\frac{\partial P_n}{\partial t} 
	+ \sum_{k = 1}^n \left[
	\frac{\partial }{\partial u_k^r} \left( \langle \dot u_k^r |u_1,\ldots,u_n\rangle 
	P_n\right) 
	+
	\frac{\partial }{\partial u_k^i}\left(\langle \dot u_k^i |u_1,\ldots,u_n\rangle 
	P_n\right) \right] = 0,
	\label{eqB3}
	\end{equation}
where we denoted $u_n^r = \mathrm{Re}\,u_n$ and $u_n^i = \mathrm{Im}\,u_n$, and $\langle \dot u_k^r |u_1,\ldots,u_n\rangle$ means the conditional average of $\dot u_k^r$ for fixed $u_1,\ldots,u_n$ and time $t$. The derivatives $\dot u_k = \dot u_k^r + i\dot u_k^i$ are given by Eq.~(\ref{eq5})  with the vanishing viscosity $\nu = 0$. Since the right-hand side of Eq.~(\ref{eq5}) depends on the two neighboring shell speeds on each side, one needs values of the speeds $u_{-1}$, $u_0$, $u_{n+1}$ and $u_{n+2}$ in order to close Eq.~(\ref{eqB3}). Since the speeds $u_{-1}$ and $u_0$ mimic the large-scale physical boundary, they should be  deterministic (recall that, in our simulations, we chose $u_{-1} = u_0 = 0$). On the contrary, the small-scale speeds $u_{n+1}$ and $u_{n+2}$ are determined by the turbulent dynamics in the inertial interval, so they are stochastic. 

Recall that every shell speed can be given as a function $u_n = u_n(\mathbf{z}_1,\ldots,\mathbf{z}_n)$ for given $u_0$ and $u_{-1}$ with the recurrence relations (\ref{eqB1b}). 
Thus, we can to do the closure of Eq.~(\ref{eqB3}) in terms of the probability distribution $\tilde{P}_n(\mathbf{z}_1,\ldots,\mathbf{z}_n)$ for the multiplier variables $\mathbf{z}_n = (w_n,\Delta_n)$, using the exact relation
	\begin{equation}
	\tilde{P}_{n+2}(\mathbf{z}_1,\ldots,\mathbf{z}_{n},\mathbf{z}_{n+1},\mathbf{z}_{n+2}) = 
	\tilde{P}_n(\mathbf{z}_1,\ldots,\mathbf{z}_n)
	P_c(\mathbf{z}_{n+1}|\mathbf{z}_{n},\mathbf{z}_{n-1},\ldots)
	P_c(\mathbf{z}_{n+2}|\mathbf{z}_{n+1},\mathbf{z}_{n},\ldots),
	\label{eqB4}
	\end{equation}
where $P_c(\mathbf{z}_n|\mathbf{z}_{n-1},\mathbf{z}_{n-2},\ldots)$ is a conditional probability. If $n$ belongs to the inertial interval, the function $P_c$ is universal, independent of $n$ and has a rapidly vanishing dependence of $\mathbf{z}_{n-j}$ for increasing $j$, as we argued in the previous section. This means that $P_c$ can be considered to be given, e.g., from numerical simulations. This provides an explicit closure, written as a continuity equation for $\tilde P_n$ similarly to Eq.~(\ref{eqB3}) as
 	\begin{equation}
	\frac{\partial \tilde{P}_n}{\partial t} 
	+ \sum_{k = 1}^n \int dw_{n+1}dw_{n+2}d\Delta_{n+1}d\Delta_{n+2} \left[
	\frac{\partial }{\partial w_n} \left(\dot w_n 
	\tilde P_{n+2}\right) 
	+\frac{\partial }{\partial \Delta_n}\left(\dot\Delta_n 
	\tilde P_{n+2}\right)\right] = 0,
	\label{eqB5}
	\end{equation}
where $\tilde{P}_{n+2}$ is given by Eq.~(\ref{eqB4}), and all derivatives $\dot w_n$ and $\dot\Delta_n$ are defined explicitly in terms of $\mathbf{z}_1,\ldots,\mathbf{z}_{n+2}$ (the formulas are given in~\cite{eyink2003gibbsian}; note a notation difference due to the factor $k_n^{1/3}$).
For the probability distribution obtained in the limit $\mathrm{Re} \to \infty$, the described purely inviscid formulation is expected to be exact in the limit of large $n$.  

We arrived at a closed formulation for the evolution of a probability distribution, which is valid for a purely inviscid flow, in the limit $\mathrm{Re} \to \infty$. Instead of the viscous regularization corresponding to finite Reynolds numbers, the inviscid continuity equation (\ref{eqB5}) features the stochastic anomaly governed by the probabilistic ``boundary condition'' (\ref{eqB4}) at small scales (large $n$). Since this condition is written in terms of random multipliers, it is natural that the stochastic component enters the flow spontaneously after the blowup when all the small-scale speeds $u_n$ get excited, see Fig.~\ref{fig2}(a). Now we are ready to verify the proposed stochastic definition of inviscid dynamics numerically.

\section{Purely inviscid turbulence: numerical tests}
\label{sec8}

We simulate the purely inviscid evolution of the probability distribution by computing numerically $10^4$ random sample solutions of the inviscid Sabra model for the shells $u_1(t),\ldots,u_n(t)$. We consider the initial time $t = 0.8$ and take the initial, boundary and forcing conditions from the viscous simulations for $\mathrm{Re} = 10^{10}$, which were described in Sections~\ref{sec4} and \ref{sec6} and will be used for comparison. Our initial time $t = 0.8$ is relatively close to the blowup and one can see in Fig.~\ref{fig2}(a) that the inertial interval is well established starting from the shell $n = 7$. Thus, we perform two different numerical tests, one with the minimum number of simulated shells $n = 7$ and the other with $n = 10$. In each simulation, the integration is carried out using the explicit second-order Adams--Bashforth method,
where the shells $u_{n+1}(t)$ and $u_{n+2}(t)$ are chosen randomly at each time step according to the universal conditional probability $P_c$ as we explain below. 

The universal conditional probability is implemented in terms of the multipliers $\mathbf{z}_k = (w_k,\Delta_k)$, which uniquely define the shell speeds, as explained in Section~\ref{sec7}.
Since correlations between $\mathbf{z}_k$ decay rapidly with a shell separation, see Fig.~\ref{fig4}(b), we limit our consideration to the four subsequent shell numbers in $P_c(\mathbf{z}_k|\mathbf{z}_{k-1},\mathbf{z}_{k-2},\mathbf{z}_{k-3})$. We determine $P_c$ numerically by using $10^4$ random samples $(\hat{\mathbf{z}}_1,\hat{\mathbf{z}}_2,\hat{\mathbf{z}}_3,\hat{\mathbf{z}}_4) = (\mathbf{z}_{k-3},\ldots,\mathbf{z}_k)$, which are taken at different times in the inertial interval for a single numerical simulation of the Sabra model. This data is computed once and stored. Then, at each time step of the integration method, given the multipliers $\mathbf{z}_{n-2}(t),\mathbf{z}_{n-1}(t),\mathbf{z}_{n}(t)$, one choses the nearest triple $\hat{\mathbf{z}}_1,\hat{\mathbf{z}}_2,\hat{\mathbf{z}}_3$ and assigns the corresponding value $\mathbf{z}_{n+1}(t) = \hat{\mathbf{z}}_4$. Next, the same procedure is applied to define $\mathbf{z}_{n+2}(t)$. These multipliers define the shell speeds $u_{n+1}(t)$ and $u_{n+2}(t)$ necessary for computing the right-hand sides of the Sabra model equations (\ref{eq5}).

\begin{figure}
\centering
\includegraphics[width = 0.7\textwidth]{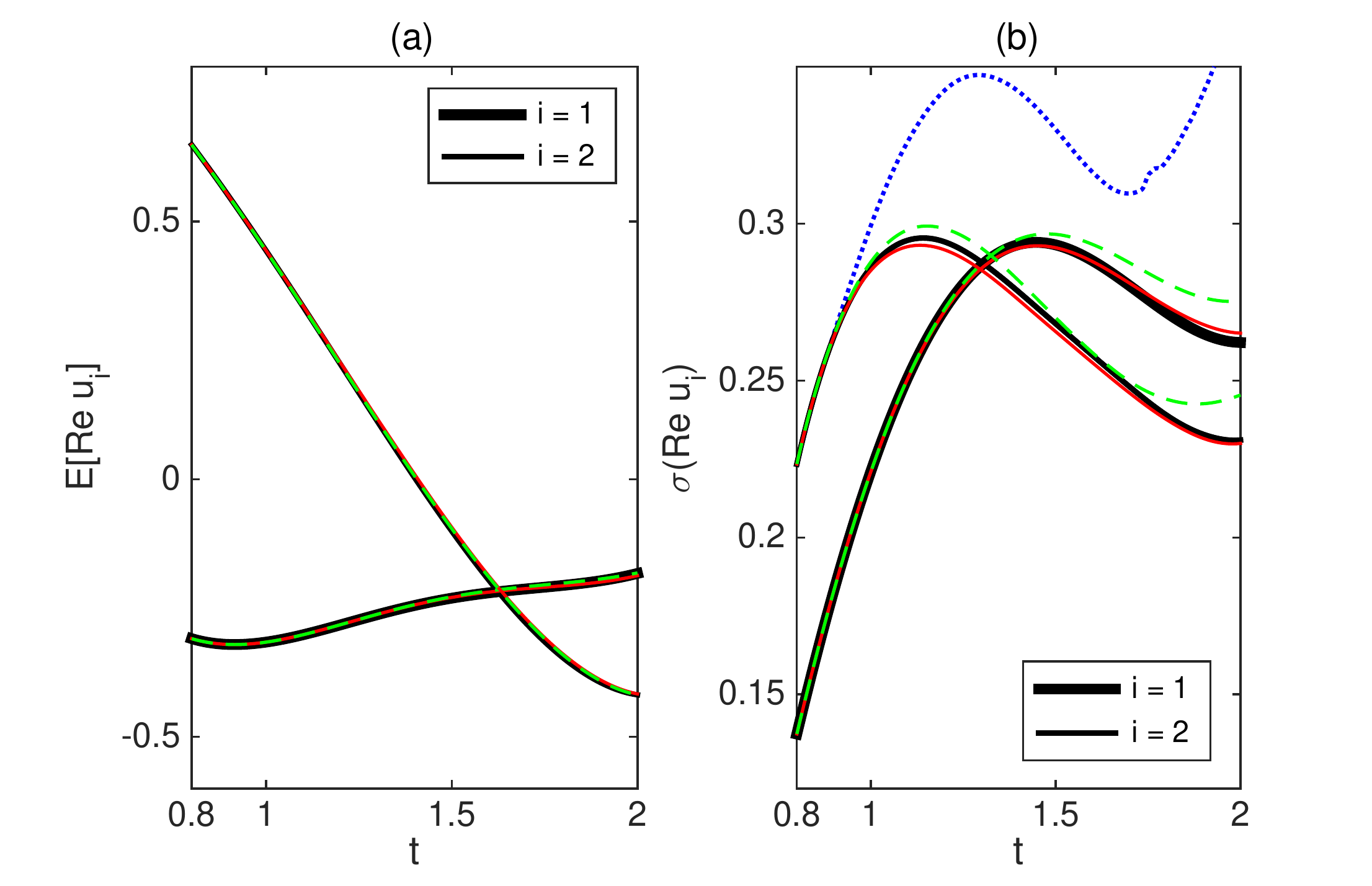}
\caption{(a) Mean values and (b) standard deviations for real parts of the large-scale shell speeds $u_1$ and $u_2$ as functions of time. Bold black curves correspond to the solution with $\mathrm{Re} = 10^{10}$ (same as in Fig.~\ref{fig1extra}). Solid red (dashed green) curves correspond to the purely inviscid stochastic evolution simulated with $n = 10$ ($n = 7$) shells. The blue dotted curve in plot (b) corresponds to a different random choice of $u_{n+1}$ and $u_{n+2}$ (for $n = 10$) highlighting the strong dependence on a stochastic small-scale boundary condition. 
}
\label{fig5}
\end{figure}

The last column in Fig.~\ref{fig1} presents the simulation results performed with a total number of $n = 7$ shells. The probability density of the second shell speed $u_2(t)$ is shown on complex plane at three different times. It is quite remarkable that with only $7$ shells, i.e., effectively with the shells of the forcing range only, the inviscid probabilistic formulation reproduces accurately the results previously obtained with very high Reynolds numbers, see the second and third columns in Fig.~\ref{fig1}. For quantitative comparison we plot in Fig.~\ref{fig5} the mean values and standard deviations of the real parts, $\mathrm{Re}\,u_1(t)$ and $\mathrm{Re}\,u_2(t)$, as functions of time. Here the bold black curves correspond to numerical simulations of Section~\ref{sec4} with the Reynolds number $\mathrm{Re} = 10^{10}$ (see also Fig.~\ref{fig1extra}), which are compared with the results of the inviscid probabilistic formulation with $n = 10$ shells (solid red lines) and $n = 7$ shells (dashed green lines). One can see that the results for the mean values are almost undistinguishable.  The standard deviations in the case $n = 7$ have a small difference, which decreases substantially for $n = 10$. 

We conclude that the obtained numerical results strongly support the theoretical construction of Section~\ref{sec7}. In order to see that the universal stochastic boundary condition represents a fine ``tuning'' of the solution in the inertial interval, we performed simulations with a different boundary condition. Namely, we considered the shells $u_{n+1}(t) = 2^{-1/3}e^{i\theta_1}u_n(t)$ and $u_{n+2}(t) = 2^{-2/3}e^{i\theta_2}u_n(t)$, which satisfy the Kolmogorov scaling and have independent uniformly distributed random phases $\theta_1$ and $\theta_2$. The resulting standard deviation of the real part  $\mathrm{Re}\,u_2(t)$ is presented in Fig.~\ref{fig5}(b) by the dotted blue curve, which rapidly diverges from the high-Reynolds-number solution.

\section{Conclusions}

The problem of turbulence for incompressible flows at large Reynolds numbers can be formulated as predicting large-scale flow through an adequate description of small-scale fluctuations. In this work, we showed that even if forcing, initial and boundary conditions are deterministic, the physically relevant flow description leads to  intrinsically stochastic dynamics. The corresponding probability distribution is well-defined at each \textit{finite time} in the limit of large Reynolds numbers if one uses a regularization term with a random viscosity. In this way, the physically relevant stochastic solutions follow in the limit of vanishing (random) viscosity, similarly to discontinuous solutions for the Burgers equation that appear in the limit of vanishing (deterministic) viscosity. We also show that the developed turbulent flow allows a purely inviscid description, which is defined for the probability distribution by means of a universal stochastic ``boundary condition'' at small scales. An important implication is that such a probability distribution can be predicted (computed) at each finite time.

We provided an extensive numerical evidence that the above description of the turbulent flow is valid for the Sabra shell model of turbulence. We argue that the stochastic anomaly is fundamentally different from the finite-dimensional chaos, where the evolution is intrinsically deterministic at any finite time. On the other hand, it is closely related to the phenomenon of spontaneous stochasticity due to non-uniqueness of Lagrangian trajectories in rough (singular) velocity fields. As we demonstrated for the Sabra model, the singular velocity appears naturally in the evolution as a result of a finite-time blowup, immediately triggering the stochastic behavior. This provides an example of a classical system whose finite-time behavior can only be described with the probabilistic terminology, a property, which is normally attributed exclusively to quantum systems. 

Our theoretical considerations are based the Kolmogorov--Obukhov theory and the concept of inverse cascade of stochasticity discussed in the literature for long time~\cite{leith1972predictability,eyink1996turbulence}. Since these ideas are valid both for the Sabra model and for the 3D Navier--Stokes equations, we conjecture that a similar formalism applies to the realistic turbulence for high Reynolds numbers. Though the existence of a finite-time blowup (that may ``turn on'' the stochasticity) remains an open problem for the inviscid flow (3D Euler equations)~\cite{gibbon2008three}, numerical simulations suggest the blowup at a physical boundary~\cite{luo2013potentially}, while nearly exponential vorticity growth is typical far from the boundary, e.g.,~\cite{brachet1992numerical,agafontsev2015}. The spontaneously stochastic  distribution for the Navier--Stokes equations can be accessed numerically by performing a large number of simulations for the same deterministic large-scale conditions and small random viscosities, which is a challenging problem due to very high requirements for computational resources. 

{\it Acknowledgments.} 
This work was supported by the CNPq (Grant No. 305519/2012-3) and FAPERJ (Pensa Rio--2014). The author is grateful to G.L. Eyink for discussion and useful comments.

\bibliography{refs}

\begin{thebibliography}{31}%
\makeatletter
\providecommand \@ifxundefined [1]{%
 \@ifx{#1\undefined}
}%
\providecommand \@ifnum [1]{%
 \ifnum #1\expandafter \@firstoftwo
 \else \expandafter \@secondoftwo
 \fi
}%
\providecommand \@ifx [1]{%
 \ifx #1\expandafter \@firstoftwo
 \else \expandafter \@secondoftwo
 \fi
}%
\providecommand \natexlab [1]{#1}%
\providecommand \enquote  [1]{``#1''}%
\providecommand \bibnamefont  [1]{#1}%
\providecommand \bibfnamefont [1]{#1}%
\providecommand \citenamefont [1]{#1}%
\providecommand \href@noop [0]{\@secondoftwo}%
\providecommand \href [0]{\begingroup \@sanitize@url \@href}%
\providecommand \@href[1]{\@@startlink{#1}\@@href}%
\providecommand \@@href[1]{\endgroup#1\@@endlink}%
\providecommand \@sanitize@url [0]{\catcode `\\12\catcode `\$12\catcode
  `\&12\catcode `\#12\catcode `\^12\catcode `\_12\catcode `\%12\relax}%
\providecommand \@@startlink[1]{}%
\providecommand \@@endlink[0]{}%
\providecommand \url  [0]{\begingroup\@sanitize@url \@url }%
\providecommand \@url [1]{\endgroup\@href {#1}{\urlprefix }}%
\providecommand \urlprefix  [0]{URL }%
\providecommand \Eprint [0]{\href }%
\providecommand \doibase [0]{http://dx.doi.org/}%
\providecommand \selectlanguage [0]{\@gobble}%
\providecommand \bibinfo  [0]{\@secondoftwo}%
\providecommand \bibfield  [0]{\@secondoftwo}%
\providecommand \translation [1]{[#1]}%
\providecommand \BibitemOpen [0]{}%
\providecommand \bibitemStop [0]{}%
\providecommand \bibitemNoStop [0]{.\EOS\space}%
\providecommand \EOS [0]{\spacefactor3000\relax}%
\providecommand \BibitemShut  [1]{\csname bibitem#1\endcsname}%
\let\auto@bib@innerbib\@empty
\bibitem [{\citenamefont {Kolmogorov}(1941)}]{kolmogorov1941local}%
  \BibitemOpen
  \bibfield  {author} {\bibinfo {author} {\bibfnamefont {A.~N.}\ \bibnamefont
  {Kolmogorov}},\ }\href@noop {} {\bibfield  {journal} {\bibinfo  {journal}
  {Dokl. Akad. Nauk SSSR}\ }\textbf {\bibinfo {volume} {30}},\ \bibinfo {pages}
  {299} (\bibinfo {year} {1941})}\BibitemShut {NoStop}%
\bibitem [{\citenamefont {Obukhov}(1941)}]{obukhov1941spectral}%
  \BibitemOpen
  \bibfield  {author} {\bibinfo {author} {\bibfnamefont {A.~M.}\ \bibnamefont
  {Obukhov}},\ }\href@noop {} {\bibfield  {journal} {\bibinfo  {journal} {Dokl.
  Akad. Nauk SSSR}\ }\textbf {\bibinfo {volume} {32}},\ \bibinfo {pages} {22}
  (\bibinfo {year} {1941})}\BibitemShut {NoStop}%
\bibitem [{\citenamefont {Scheffer}(1993)}]{scheffer1993inviscid}%
  \BibitemOpen
  \bibfield  {author} {\bibinfo {author} {\bibfnamefont {V.}~\bibnamefont
  {Scheffer}},\ }\href@noop {} {\bibfield  {journal} {\bibinfo  {journal} {J.
  Geom. Anal.}\ }\textbf {\bibinfo {volume} {3}},\ \bibinfo {pages} {343}
  (\bibinfo {year} {1993})}\BibitemShut {NoStop}%
\bibitem [{\citenamefont {Shnirelman}(1997)}]{shnirelman1997nonuniqueness}%
  \BibitemOpen
  \bibfield  {author} {\bibinfo {author} {\bibfnamefont {A.}~\bibnamefont
  {Shnirelman}},\ }\href@noop {} {\bibfield  {journal} {\bibinfo  {journal}
  {Commun. Pure Appl. Math.}\ }\textbf {\bibinfo {volume} {50}},\ \bibinfo
  {pages} {1261} (\bibinfo {year} {1997})}\BibitemShut {NoStop}%
\bibitem [{\citenamefont {Bardos}\ and\ \citenamefont
  {Titi}(2013)}]{bardos2013mathematics}%
  \BibitemOpen
  \bibfield  {author} {\bibinfo {author} {\bibfnamefont {C.~W.}\ \bibnamefont
  {Bardos}}\ and\ \bibinfo {author} {\bibfnamefont {E.~S.}\ \bibnamefont
  {Titi}},\ }\href@noop {} {\bibfield  {journal} {\bibinfo  {journal} {J.
  Turbul.}\ }\textbf {\bibinfo {volume} {14}},\ \bibinfo {pages} {42} (\bibinfo
  {year} {2013})}\BibitemShut {NoStop}%
\bibitem [{\citenamefont {Frisch}(1999)}]{frisch1999turbulence}%
  \BibitemOpen
  \bibfield  {author} {\bibinfo {author} {\bibfnamefont {U.}~\bibnamefont
  {Frisch}},\ }\href@noop {} {\emph {\bibinfo {title} {{Turbulence: the legacy
  of A.N.~Kolmogorov}}}}\ (\bibinfo  {publisher} {Cambridge University Press},\
  \bibinfo {year} {1999})\BibitemShut {NoStop}%
\bibitem [{\citenamefont {Bernard}\ \emph {et~al.}(1998)\citenamefont
  {Bernard}, \citenamefont {Gawedzki},\ and\ \citenamefont
  {Kupiainen}}]{bernard1998slow}%
  \BibitemOpen
  \bibfield  {author} {\bibinfo {author} {\bibfnamefont {D.}~\bibnamefont
  {Bernard}}, \bibinfo {author} {\bibfnamefont {K.}~\bibnamefont {Gawedzki}}, \
  and\ \bibinfo {author} {\bibfnamefont {A.}~\bibnamefont {Kupiainen}},\
  }\href@noop {} {\bibfield  {journal} {\bibinfo  {journal} {J. Stat. Phys.}\
  }\textbf {\bibinfo {volume} {90}},\ \bibinfo {pages} {519} (\bibinfo {year}
  {1998})}\BibitemShut {NoStop}%
\bibitem [{\citenamefont {E}\ and\ \citenamefont
  {Vanden-Eijnden}(2000)}]{eijnden2000generalized}%
  \BibitemOpen
  \bibfield  {author} {\bibinfo {author} {\bibfnamefont {W.}~\bibnamefont {E}}\
  and\ \bibinfo {author} {\bibfnamefont {E.}~\bibnamefont {Vanden-Eijnden}},\
  }\href@noop {} {\bibfield  {journal} {\bibinfo  {journal} {Proc. Natl. Acad.
  Sci. USA}\ }\textbf {\bibinfo {volume} {97}},\ \bibinfo {pages} {8200}
  (\bibinfo {year} {2000})}\BibitemShut {NoStop}%
\bibitem [{\citenamefont {Falkovich}\ \emph {et~al.}(2001)\citenamefont
  {Falkovich}, \citenamefont {Gaw\ifmmode~\mbox{\c{e}}\else \c{e}\fi{}dzki},\
  and\ \citenamefont {Vergassola}}]{falkovich2001particles}%
  \BibitemOpen
  \bibfield  {author} {\bibinfo {author} {\bibfnamefont {G.}~\bibnamefont
  {Falkovich}}, \bibinfo {author} {\bibfnamefont {K.}~\bibnamefont
  {Gaw\ifmmode~\mbox{\c{e}}\else \c{e}\fi{}dzki}}, \ and\ \bibinfo {author}
  {\bibfnamefont {M.}~\bibnamefont {Vergassola}},\ }\href@noop {} {\bibfield
  {journal} {\bibinfo  {journal} {Rev. Mod. Phys.}\ }\textbf {\bibinfo {volume}
  {73}},\ \bibinfo {pages} {913} (\bibinfo {year} {2001})}\BibitemShut
  {NoStop}%
\bibitem [{\citenamefont {Kupiainen}(2003)}]{kupiainen2003nondeterministic}%
  \BibitemOpen
  \bibfield  {author} {\bibinfo {author} {\bibfnamefont {A.}~\bibnamefont
  {Kupiainen}},\ }\href@noop {} {\bibfield  {journal} {\bibinfo  {journal}
  {Ann. Henri Poincar{\'e}}\ }\textbf {\bibinfo {volume} {4}},\ \bibinfo
  {pages} {S713} (\bibinfo {year} {2003})}\BibitemShut {NoStop}%
\bibitem [{\citenamefont {Leith}\ and\ \citenamefont
  {Kraichnan}(1972)}]{leith1972predictability}%
  \BibitemOpen
  \bibfield  {author} {\bibinfo {author} {\bibfnamefont {C.~E.}\ \bibnamefont
  {Leith}}\ and\ \bibinfo {author} {\bibfnamefont {R.~H.}\ \bibnamefont
  {Kraichnan}},\ }\href@noop {} {\bibfield  {journal} {\bibinfo  {journal} {J.
  Atmos. Sci.}\ }\textbf {\bibinfo {volume} {29}},\ \bibinfo {pages} {1041}
  (\bibinfo {year} {1972})}\BibitemShut {NoStop}%
\bibitem [{\citenamefont {Ruelle}(1979)}]{ruelle1979microscopic}%
  \BibitemOpen
  \bibfield  {author} {\bibinfo {author} {\bibfnamefont {D.}~\bibnamefont
  {Ruelle}},\ }\href@noop {} {\bibfield  {journal} {\bibinfo  {journal} {Phys.
  Lett. A}\ }\textbf {\bibinfo {volume} {72}},\ \bibinfo {pages} {81} (\bibinfo
  {year} {1979})}\BibitemShut {NoStop}%
\bibitem [{\citenamefont {Eyink}(1996)}]{eyink1996turbulence}%
  \BibitemOpen
  \bibfield  {author} {\bibinfo {author} {\bibfnamefont {G.~L.}\ \bibnamefont
  {Eyink}},\ }\href@noop {} {\bibfield  {journal} {\bibinfo  {journal} {J.
  Stat. Phys.}\ }\textbf {\bibinfo {volume} {83}},\ \bibinfo {pages} {955}
  (\bibinfo {year} {1996})}\BibitemShut {NoStop}%
\bibitem [{\citenamefont {Eyink}(2014)}]{eyink2014}%
  \BibitemOpen
  \bibfield  {author} {\bibinfo {author} {\bibfnamefont {G.~L.}\ \bibnamefont
  {Eyink}},\ }\href@noop {} {\enquote {\bibinfo {title} {Mathematical analysis
  of turbulence},}\ } (\bibinfo {year} {2014}),\ \bibinfo {note} {{Lectures at
  the Mathematics of Turbulence Tutorials at IPAM, UCLA, Los Angeles,
  CA}}\BibitemShut {NoStop}%
\bibitem [{\citenamefont {Landau}\ and\ \citenamefont
  {Lifshitz}(2013)}]{landau2013fluid}%
  \BibitemOpen
  \bibfield  {author} {\bibinfo {author} {\bibfnamefont {L.~D.}\ \bibnamefont
  {Landau}}\ and\ \bibinfo {author} {\bibfnamefont {E.~M.}\ \bibnamefont
  {Lifshitz}},\ }\href@noop {} {\emph {\bibinfo {title} {{Fluid Mechanics.
  Course of Theoretical Physics}}}},\ Vol.~\bibinfo {volume} {6}\ (\bibinfo
  {publisher} {Elsevier},\ \bibinfo {year} {2013})\BibitemShut {NoStop}%
\bibitem [{\citenamefont {Onsager}(1949)}]{onsager1949statistical}%
  \BibitemOpen
  \bibfield  {author} {\bibinfo {author} {\bibfnamefont {L.}~\bibnamefont
  {Onsager}},\ }\href@noop {} {\bibfield  {journal} {\bibinfo  {journal} {Nuovo
  Cimento}\ }\textbf {\bibinfo {volume} {6}},\ \bibinfo {pages} {279} (\bibinfo
  {year} {1949})}\BibitemShut {NoStop}%
\bibitem [{\citenamefont {Eyink}\ and\ \citenamefont
  {Sreenivasan}(2006)}]{eyink2006onsager}%
  \BibitemOpen
  \bibfield  {author} {\bibinfo {author} {\bibfnamefont {G.~L.}\ \bibnamefont
  {Eyink}}\ and\ \bibinfo {author} {\bibfnamefont {K.~R.}\ \bibnamefont
  {Sreenivasan}},\ }\href@noop {} {\bibfield  {journal} {\bibinfo  {journal}
  {Rev. Modern Phys.}\ }\textbf {\bibinfo {volume} {78}},\ \bibinfo {pages}
  {87} (\bibinfo {year} {2006})}\BibitemShut {NoStop}%
\bibitem [{\citenamefont {L'vov}\ \emph {et~al.}(1998)\citenamefont {L'vov},
  \citenamefont {Podivilov}, \citenamefont {Pomyalov}, \citenamefont
  {Procaccia},\ and\ \citenamefont {Vandembroucq}}]{l1998improved}%
  \BibitemOpen
  \bibfield  {author} {\bibinfo {author} {\bibfnamefont {V.~S.}\ \bibnamefont
  {L'vov}}, \bibinfo {author} {\bibfnamefont {E.}~\bibnamefont {Podivilov}},
  \bibinfo {author} {\bibfnamefont {A.}~\bibnamefont {Pomyalov}}, \bibinfo
  {author} {\bibfnamefont {I.}~\bibnamefont {Procaccia}}, \ and\ \bibinfo
  {author} {\bibfnamefont {D.}~\bibnamefont {Vandembroucq}},\ }\href@noop {}
  {\bibfield  {journal} {\bibinfo  {journal} {Phys. Rev. E}\ }\textbf {\bibinfo
  {volume} {58}},\ \bibinfo {pages} {1811} (\bibinfo {year}
  {1998})}\BibitemShut {NoStop}%
\bibitem [{\citenamefont {Dombre}\ and\ \citenamefont
  {Gilson}(1998)}]{dombre1998intermittency}%
  \BibitemOpen
  \bibfield  {author} {\bibinfo {author} {\bibfnamefont {T.}~\bibnamefont
  {Dombre}}\ and\ \bibinfo {author} {\bibfnamefont {J.-L.}\ \bibnamefont
  {Gilson}},\ }\href@noop {} {\bibfield  {journal} {\bibinfo  {journal}
  {Physica D}\ }\textbf {\bibinfo {volume} {111}},\ \bibinfo {pages} {265}
  (\bibinfo {year} {1998})}\BibitemShut {NoStop}%
\bibitem [{\citenamefont {Mailybaev}(2012)}]{mailybaev2012renormalization}%
  \BibitemOpen
  \bibfield  {author} {\bibinfo {author} {\bibfnamefont {A.~A.}\ \bibnamefont
  {Mailybaev}},\ }\href@noop {} {\bibfield  {journal} {\bibinfo  {journal}
  {Phys. Rev. E}\ }\textbf {\bibinfo {volume} {85}},\ \bibinfo {pages} {066317}
  (\bibinfo {year} {2012})}\BibitemShut {NoStop}%
\bibitem [{\citenamefont {Constantin}\ \emph {et~al.}(2007)\citenamefont
  {Constantin}, \citenamefont {Levant},\ and\ \citenamefont
  {Titi}}]{constantin2007regularity}%
  \BibitemOpen
  \bibfield  {author} {\bibinfo {author} {\bibfnamefont {P.}~\bibnamefont
  {Constantin}}, \bibinfo {author} {\bibfnamefont {B.}~\bibnamefont {Levant}},
  \ and\ \bibinfo {author} {\bibfnamefont {E.~S.}\ \bibnamefont {Titi}},\
  }\href@noop {} {\bibfield  {journal} {\bibinfo  {journal} {Phys. Rev. E}\
  }\textbf {\bibinfo {volume} {75}},\ \bibinfo {pages} {016304} (\bibinfo
  {year} {2007})}\BibitemShut {NoStop}%
\bibitem [{\citenamefont {Mailybaev}(2015)}]{mailybaev2015spontaneous}%
  \BibitemOpen
  \bibfield  {author} {\bibinfo {author} {\bibfnamefont {A.~A.}\ \bibnamefont
  {Mailybaev}},\ }\href@noop {} {\bibfield  {journal} {\bibinfo  {journal}
  {arXiv:1504.00575}\ } (\bibinfo {year} {2015})}\BibitemShut {NoStop}%
\bibitem [{\citenamefont {Biferale}(2003)}]{biferale2003shell}%
  \BibitemOpen
  \bibfield  {author} {\bibinfo {author} {\bibfnamefont {L.}~\bibnamefont
  {Biferale}},\ }\href@noop {} {\bibfield  {journal} {\bibinfo  {journal}
  {Annu. Rev. Fluid Mech.}\ }\textbf {\bibinfo {volume} {35}},\ \bibinfo
  {pages} {441} (\bibinfo {year} {2003})}\BibitemShut {NoStop}%
\bibitem [{\citenamefont {Kolmogorov}(1962)}]{kolmogorov1962refinement}%
  \BibitemOpen
  \bibfield  {author} {\bibinfo {author} {\bibfnamefont {A.~N.}\ \bibnamefont
  {Kolmogorov}},\ }\href@noop {} {\bibfield  {journal} {\bibinfo  {journal} {J.
  Fluid Mech.}\ }\textbf {\bibinfo {volume} {13}},\ \bibinfo {pages} {82}
  (\bibinfo {year} {1962})}\BibitemShut {NoStop}%
\bibitem [{\citenamefont {Chen}\ \emph {et~al.}(2003)\citenamefont {Chen},
  \citenamefont {Chen}, \citenamefont {Eyink},\ and\ \citenamefont
  {Sreenivasan}}]{chen2003kolmogorov}%
  \BibitemOpen
  \bibfield  {author} {\bibinfo {author} {\bibfnamefont {Q.}~\bibnamefont
  {Chen}}, \bibinfo {author} {\bibfnamefont {S.}~\bibnamefont {Chen}}, \bibinfo
  {author} {\bibfnamefont {G.~L.}\ \bibnamefont {Eyink}}, \ and\ \bibinfo
  {author} {\bibfnamefont {K.~R.}\ \bibnamefont {Sreenivasan}},\ }\href@noop {}
  {\bibfield  {journal} {\bibinfo  {journal} {Phys. Rev. Lett.}\ }\textbf
  {\bibinfo {volume} {90}},\ \bibinfo {pages} {254501} (\bibinfo {year}
  {2003})}\BibitemShut {NoStop}%
\bibitem [{\citenamefont {Benzi}\ \emph {et~al.}(1993)\citenamefont {Benzi},
  \citenamefont {Biferale},\ and\ \citenamefont
  {Parisi}}]{benzi1993intermittency}%
  \BibitemOpen
  \bibfield  {author} {\bibinfo {author} {\bibfnamefont {R.}~\bibnamefont
  {Benzi}}, \bibinfo {author} {\bibfnamefont {L.}~\bibnamefont {Biferale}}, \
  and\ \bibinfo {author} {\bibfnamefont {G.}~\bibnamefont {Parisi}},\
  }\href@noop {} {\bibfield  {journal} {\bibinfo  {journal} {Physica D}\
  }\textbf {\bibinfo {volume} {65}},\ \bibinfo {pages} {163} (\bibinfo {year}
  {1993})}\BibitemShut {NoStop}%
\bibitem [{\citenamefont {Eyink}\ \emph {et~al.}(2003)\citenamefont {Eyink},
  \citenamefont {Chen},\ and\ \citenamefont {Chen}}]{eyink2003gibbsian}%
  \BibitemOpen
  \bibfield  {author} {\bibinfo {author} {\bibfnamefont {G.~L.}\ \bibnamefont
  {Eyink}}, \bibinfo {author} {\bibfnamefont {S.}~\bibnamefont {Chen}}, \ and\
  \bibinfo {author} {\bibfnamefont {Q.}~\bibnamefont {Chen}},\ }\href@noop {}
  {\bibfield  {journal} {\bibinfo  {journal} {J. Stat. Phys.}\ }\textbf
  {\bibinfo {volume} {113}},\ \bibinfo {pages} {719} (\bibinfo {year}
  {2003})}\BibitemShut {NoStop}%
\bibitem [{\citenamefont {Gibbon}(2008)}]{gibbon2008three}%
  \BibitemOpen
  \bibfield  {author} {\bibinfo {author} {\bibfnamefont {J.~D.}\ \bibnamefont
  {Gibbon}},\ }\href@noop {} {\bibfield  {journal} {\bibinfo  {journal}
  {Physica D}\ }\textbf {\bibinfo {volume} {237}},\ \bibinfo {pages} {1894}
  (\bibinfo {year} {2008})}\BibitemShut {NoStop}%
\bibitem [{\citenamefont {Luo}\ and\ \citenamefont
  {Hou}(2014)}]{luo2013potentially}%
  \BibitemOpen
  \bibfield  {author} {\bibinfo {author} {\bibfnamefont {G.}~\bibnamefont
  {Luo}}\ and\ \bibinfo {author} {\bibfnamefont {T.~Y.}\ \bibnamefont {Hou}},\
  }\href@noop {} {\bibfield  {journal} {\bibinfo  {journal} {PNAS}\ }\textbf
  {\bibinfo {volume} {111}},\ \bibinfo {pages} {12968} (\bibinfo {year}
  {2014})}\BibitemShut {NoStop}%
\bibitem [{\citenamefont {Brachet}\ \emph {et~al.}(1992)\citenamefont
  {Brachet}, \citenamefont {Meneguzzi}, \citenamefont {Vincent}, \citenamefont
  {Politano},\ and\ \citenamefont {Sulem}}]{brachet1992numerical}%
  \BibitemOpen
  \bibfield  {author} {\bibinfo {author} {\bibfnamefont {M.~E.}\ \bibnamefont
  {Brachet}}, \bibinfo {author} {\bibfnamefont {M.}~\bibnamefont {Meneguzzi}},
  \bibinfo {author} {\bibfnamefont {A.}~\bibnamefont {Vincent}}, \bibinfo
  {author} {\bibfnamefont {H.}~\bibnamefont {Politano}}, \ and\ \bibinfo
  {author} {\bibfnamefont {P.~L.}\ \bibnamefont {Sulem}},\ }\href@noop {}
  {\bibfield  {journal} {\bibinfo  {journal} {Phys. Fluids A}\ }\textbf
  {\bibinfo {volume} {4}},\ \bibinfo {pages} {2845} (\bibinfo {year}
  {1992})}\BibitemShut {NoStop}%
\bibitem [{\citenamefont {Agafontsev}\ \emph {et~al.}(2015)\citenamefont
  {Agafontsev}, \citenamefont {Kuznetsov},\ and\ \citenamefont
  {Mailybaev}}]{agafontsev2015}%
  \BibitemOpen
  \bibfield  {author} {\bibinfo {author} {\bibfnamefont {D.~S.}\ \bibnamefont
  {Agafontsev}}, \bibinfo {author} {\bibfnamefont {E.~A.}\ \bibnamefont
  {Kuznetsov}}, \ and\ \bibinfo {author} {\bibfnamefont {A.~A.}\ \bibnamefont
  {Mailybaev}},\ }\href@noop {} {\bibfield  {journal} {\bibinfo  {journal}
  {Phys. Fluids}\ }\textbf {\bibinfo {volume} {27}},\ \bibinfo {pages} {085102}
  (\bibinfo {year} {2015})}\BibitemShut {NoStop}%
\end{thebibliography}%

\end{document}